\documentclass[manuscript]{aastex63}

\usepackage{booktabs}
\received{May 2, 2020}
\revised{July 6, 2020}
\accepted{July 6, 2020}
\submitjournal{ApJS}

\shorttitle{Rotation periods in 1000 \textit{TESS} objects of interest}
\shortauthors{Canto Martins et al.}

\begin{document}

\title{A search for rotation periods in 1000 \textit{TESS} objects of interest}

\correspondingauthor{Bruno L. Canto Martins}
\email{brunocanto@fisica.ufrn.br}

\author[0000-0001-5578-7400]{B. L. Canto Martins}
\affiliation{Departamento de F\'isica Te\'orica e Experimental, Universidade Federal do Rio Grande do Norte, Campus Universit\'ario, Natal, RN, 59072-970, Brazil}
\author[0000-0002-2023-7641]{R. L. Gomes}
\affiliation{Departamento de F\'isica Te\'orica e Experimental, Universidade Federal do Rio Grande do Norte, Campus Universit\'ario, Natal, RN, 59072-970, Brazil}
\author[0000-0002-2425-801X]{Y. S. Messias}
\affiliation{Departamento de F\'isica Te\'orica e Experimental, Universidade Federal do Rio Grande do Norte, Campus Universit\'ario, Natal, RN, 59072-970, Brazil}
\author[0000-0002-2702-231X]{S. R. de Lira}
\affiliation{Departamento de F\'isica Te\'orica e Experimental, Universidade Federal do Rio Grande do Norte, Campus Universit\'ario, Natal, RN, 59072-970, Brazil}
\author[0000-0001-5845-947X]{I. C. Le\~ao}
\affiliation{Departamento de F\'isica Te\'orica e Experimental, Universidade Federal do Rio Grande do Norte, Campus Universit\'ario, Natal, RN, 59072-970, Brazil}
\author[0000-0002-3817-6402]{L. A. Almeida}
\affiliation{Departamento de F\'isica, Universidade do Estado do Rio Grande do Norte, Campus Universit\'ario Central, Mossor\'o, RN, 59610-090, Brazil}
\author[0000-0002-5404-8451]{M. A. Teixeira}
\affiliation{Departamento de F\'isica Te\'orica e Experimental, Universidade Federal do Rio Grande do Norte, Campus Universit\'ario, Natal, RN, 59072-970, Brazil}
\author[0000-0002-5102-0154]{M. L. das Chagas}
\affiliation{Faculdade de F\'isica/ Instituto de Ci\^encias Exatas, Universidade Federal do Sul e Sudeste do Par\'a, 68505-080, Marab\'a PA, Brazil}
\author[0000-0002-1272-524X]{J. P. Bravo}
\affiliation{Departamento de F\'isica Te\'orica e Experimental, Universidade Federal do Rio Grande do Norte, Campus Universit\'ario, Natal, RN, 59072-970, Brazil}
\author[0000-0001-5576-8365]{A. Bewketu Belete}
\affiliation{Departamento de F\'isica Te\'orica e Experimental, Universidade Federal do Rio Grande do Norte, Campus Universit\'ario, Natal, RN, 59072-970, Brazil}
\author[0000-0001-8218-1586]{J. R. De Medeiros}
\affiliation{Departamento de F\'isica Te\'orica e Experimental, Universidade Federal do Rio Grande do Norte, Campus Universit\'ario, Natal, RN, 59072-970, Brazil}




\begin{abstract}

The high quality light curves from the Transiting Exoplanet Survey Satellite (TESS) represent a unique laboratory for the study of stellar rotation, a fundamental observable driving stellar and planetary evolution, including planetary atmospheres and impacting on habitability conditions and the genesis of life around stars. As of April 14th 2020, this mission delivered public light curves for 1000 TESS Objects of Interest (TOIs), observed with 2 minute cadence during the first 20 months of the mission. Here, we present a search for rotation signatures in these TOIs, using Fast Fourier Transform, Lomb-Scargle, and wavelet techniques, accompanied by a rigorous visual inspection. This effort revealed 163 targets with rotation signatures, 131 of which present unambiguous rotation periods ranging from 0.321 and 13.219 days, whereas 32 of them present dubious rotation periodicities. One hundred and nine of these stars show flux fluctuations whose root-cause is not clearly identified. For 714 TOIs, the light curves show a noisy behavior, corresponding to typically low-amplitude signals. Our analysis has also revealed 10 TOI stars with pulsation periodicities ranging from 0.049 to 2.995 days and four eclipsing binaries. With upcoming TESS data releases, our periodicity analysis will be expanded to almost all TOI stars, thereby contributing in defining criteria for follow-up strategy itself, and the study of star-planet interactions, surface dynamic of host stars and habitability conditions in planets, among other aspects. In this context, a living catalog is maintained on the Filtergraph visualization portal at the URL  \url{https://filtergraph.com/tess_rotation_tois }.

\end{abstract}


\keywords{stars: rotation - stars: variables: general - stars: planetary systems - techniques: photometric}


\section{Introduction} \label{sec:intro}

Photometric space missions are revolutionizing our understanding of stellar periodicities, revealing a new view of the variability of stars in different regions of the HR Diagram. Thanks to the photometric observations carried out by the CoRoT (Baglin et al. 2009) and {\em Kepler} (Borucki et al. 2010) missions, different studies have revealed new insights on the rotation of main-sequence stars (e.g.; De Medeiros et al. 2013; Nielsen et al. 2013; Walkowicz \& Basri 2013; McQuillan, Mazeh \& Aigrain 2014; Le\~ao et al. 2015; Paz-Chinch\'on et al. 2015; Davenport 2017; Reinhold \& Hekker 2020), as well as in advanced stages of stellar evolution (e.g. Mosser et al. 2012; Van Saders \& Pinsonneault 2013; De Medeiros et al. 2013; Costa et al. 2015). These works have shown that rotation is a major constraint in the study of the angular momentum, including the angular momentum transport from core to surface and expanding envelope in stars. The normalcy of the Sun's rotation with respect to the main-sequence stars with surface physical parameters close to solar values (De Freitas et al. 2013; Le\~ao et al. 2015) and a bimodality in the rotation period distribution for M dwarf (McQuillan et al. 2013), K dwarf (McQuillan et al. 2014), and main-sequence stars with effective temperature above 5000~K (Davenport 2017) have also emerged from data acquired by the referred space missions. In addition, the observations of photometric modulation are also revealing traces of rotation in white dwarf stars, one of the scarce remaining clues of physics of the formation process of these stars (Maoz 2015, Kawaler 2015, de Lira et al. 2019). 

The Transiting Exoplanet Survey Satellite (TESS) space mission (Ricker et al. 2015), launched into space in April 2018, is performing a 2-year nearly all-sky survey, during which differential time-series photometry are being acquired for hundreds of thousands of stars. Although the primary goal of TESS is to search for terrestrial planets transiting nearby bright stars, the large number of observed targets enables the study of other astrophysical phenomena, including stellar periodicities. For instance, first results based on TESS observations have revealed rapidly rotating M dwarfs with periods less than 1 day (Zhan et al. 2019), rotational and pulsation variability of magnetic chemically peculiar A-type stars (Cunha et al. 2019), and the identification of flares in GKM-type stars (Howard et al. 2019;  G\"unther et al. 2019; 2020; Tu et al. 2020; Doyle et al. 2020).

The stellar environments in and around stars hosting planets are complex and unique laboratories for the understanding of the relation between the stars and orbiting companions. Much of the information about this interaction is encoded within their different variability phenomena, including rotation, pulsation and flares. Indeed, rotation is a paramount parameter driving the stellar evolution, playing also a major role in planetary evolution and habitability. The great significance of rotation is revealed by paralleling its role in the Solar System evolution, controlling the Sun's different transient phenomena, including radiative energy, the plasma outflow, shock waves, high-energy particle events during flares, and coronal mass ejections, which are key ingredients in the formation and atmospheric evolution of the planets including the terrestrial biosphere (e.g., Lundin et al. 2007; Lammer et al. 2012). In this context, the era of exoplanet transit surveys offers a unique possibility to the study of the rotation of stars hosting planets, thanks to the detection of quasi-periodic brightness variations in the photometric time series, caused by magnetically active regions crossing recurrently the visible hemisphere as the stars rotate (e.g., Irwin et al. 2009; Mc Quillan et al. 2014; Paz-Chinch\'on et al. 2015). Deriving the rotation period for large samples of stars hosting planets has been a long-standing goal in stellar astronomy, with the potential to shed light on evolution of the angular-momentum of stars and their planetary system and to understand how magnetic features affect exoplanet parameters. For instance, intensive studies of the physical properties of the planets and their parent stars, including a possible star-planet interaction, have been conducted (Canto Martins et al. 2011; Miller et al. 2015; Viswanath et al. 2020). In addition, the advance in the knowledge of the rotation period of stars is also important in the support of exoplanet search, because stellar rotation may act itself on both photometric and spectroscopic data, preventing the detection and characterization of planets with orbital period near the stellar rotation period or its harmonics.

In this work we report a search for periodicities in the first 1000 TESS Objects of Interest (TOIs), mostly focused in the identification of rotation signatures, on the base of wavelet, Fast Fourier Transform and Lomb-Scargle analyses. Indeed, the philosophy of this effort is to offer for the exoplanet community, exploring TESS observations, a diagnostic of the presence of rotation phenomena in the TOI stars. As highlighted above, this work could provide valuable information to answer a large number of questions, including follow-up strategy itself, star-planet interactions, surface dynamic of host stars and habitability conditions in planets. The paper is organized as follows. Section \ref{observation} presents the data set used in our study and discusses the analysis procedure applied in the search for variability. Section \ref{sec:results} provides the main results. A summary is presented in Section \ref{summary}. 

\section{Stellar Sample and Observational Data} \label{observation}

As underlined in the previous Section, TESS is an ongoing NASA photometric space mission and its main goal is the search for exoplanets by using the photometric transit method. In two years, the mission plan is to cover almost the entire sky by monitoring 26 segments (or sectors) of $90^o\times24^o$, each one during 27 days. In the first and second years, the mission will complete the survey of the southern and northern ecliptic hemispheres, respectively. At higher ecliptic latitudes, there are overlap regions among the sectors where the targets can be observed for 54, 81, 108, 189, and 351 days. For a detailed description of the TESS mission see Ricker et al. (2015).

For the present purpose, we selected the first 1000 TESS Objects of Interest (TOIs) to perform a global search for periodicities using different procedures. TESS mission provides photometric data at two different cadences (2 and 30 minutes) with a time baseline from 27 days to 351 days, depending on sector overlap. While the 2-minute cadence data, also known as Target Pixel (TP) files, are available for a subset of targets, the entire CCDs, called full-frame images (FFIs), are binned on-board every 30 minutes and available via internet\footnote{\url{http://archive.stsci.edu/tess}}.  

The TESS light curves (LCs) were automatically reduced and corrected for common instrumental systematics by the TESS data processing pipeline\footnote{\url{https://heasarc.gsfc.nasa.gov/docs/tess/pipeline.html}} (Jenkins et al. 2016). The TESS pipeline is based on that used by the {\em Kepler} Mission with further improvements. The data reduction performed by TESS is done using simple aperture photometry (SAP) on each TP files. The LCs for all targets are created and stored in arrays of fluxes. Subsequent detrend are applied to the LCs using the cotrending basis vectors, which represent the set of systematic trends present in the data for each CCD in each sector, and stored in other arrays called pre-search data conditioning (PCDSAP). In this study, we are using the 2-minute cadence PDCSAP data retrieved via Space Telescope at Science Institute Webpage\footnote{\url{http://archive.stsci.edu/tess/bulk_downloads/bulk_downloads_ffi-tp-lc-dv.html}}.

While the detection of periodicities in LCs is straightforward, their interpretation in terms of the root-causes is far a challenging task. Indeed, the detection threshold for periodicity it depends on star brightness, time span of observations, and the final cleaning of the LCs, varying thus from star to star. In this sense, to avoid possible distortions in the signature of periodicities, we have performed an additional treatment of outlier removal and instrumental trend correction for the LCs following the procedure by De Medeiros et al. (2013) and Paz-Chinch\'on et al. (2015). We performed such a treatment when needed plus a removal of transits in a similar way to that described in Paz-Chinch\'on et al. (2015). The reader is referred to those authors for a complete explanation on this post-treatment and data analysis, which is summarized below.

In summary, our post-treatment consisted of removing eventual {\em flare-like}\footnote{A noticeable flux bump typically of few hours, whose physical or instrumental origin was inspected separately.} signatures from the PDCSAP LCs, as well as the known planetary transits based on the TOI catalog\footnote{\url{https://tess.mit.edu/toi-releases/}}. Nevertheless, those features were analyzed separately for identification of physical flares and binarity. A few jumps were corrected based on De Medeiros et al. (2013) and B\'anyai et al. (2013), by taking a linear fit and extrapolation of user-defined boxes before and after each jump. Individual LCs of each TESS sector were then detrended with third-order polynomial fits. This step is basically a high-pass filter that helps in suppressing long-term trends usually associated to instrumental systematics (e.g., Smith et al. 2012; Basri et al. 2011). Of course, such a filter may also suppress long-period physical variabilities, but in the present study such periods would be longer than the typical 28-day time span of the TESS sectors, namely a technical limit for period determination (e.g., G\"unther et al. 2020). Finally, removal of outliers was performed by excluding any flux measurement greater than 3.5$\times$ the standard deviation of the detrended LCs. In addition, individual LCs that were overlapped in multiple sectors were combined to produce a single long-term time series for each object. These steps produced rather clean LCs without transits (or flares) that allowed inspection of stellar variations such as rotational modulation.

\subsection{Identifying periodicities} 

The post-processed LCs were analyzed by using three different periodicity analysis techniques, namely (i) Lomb-Scargle periodograms (e.g., Scargle 1982; Horne \& Baliunas 1986; Press \& Rybicki 1989), (ii) Fast Fourier Transform (FFT) (see Zhan et al. 2019 for details) and (iii) wavelet analysis (Grossmann \& Morlet 1984). In fact, we consider these three methods to identify consistent periodicities. In general, these procedures can provide additional information to a visual inspection of the LCs. It is common that the peak powers in the power spectrum (or periodogram) of a method do not follow the same sequence of another method. Even in those cases, we interpret periodicities as far as they are revealed by different methods, and given that they are statistically confident. In particular, the Lomb-Scargle method is useful for validating periods according to their false alarm probabilities (FAPs; see Horne \& Baliunas 1986), whereas the wavelet maps strongly help to interpret morphological nuances of periodic signatures (e.g., Bravo et al. 2014). We have considered a broad frequency range of 0.01--10.0 d$^{-1}$ to search for rotation and pulsation signatures. Those three methods are described shortly in Sects. 2.2, 2.3, and 2.4. The periods given in our catalog are those from the peaks of wavelet global spectra and their errors are computed using Eq. (2) of Lamm et al. (2004), being typically around 5\%.  

Different authors have described, in detail, rotational modulation as being a semi-sinusoidal variability associated to the dynamic behavior of star spots (e.g.: Basri \& Nguyen 2018; Basri 2018; De Medeiros et al. 2013; Lanza et al. 2003, 2007). In short, that signature is characterized by semi-regular flux variations that use to be multi-sinusoidal, most commonly showing single or double dips per rotation cycle. The double-dip signature has been traditionally interpreted following a simplistic view of being caused by two main spots at opposite hemispheres (e.g.: Donnelly \& Puga 1990; Lanza et al. 2009; Walkowicz et al. 2013; De Medeiros et al. 2013). However, based on Basri \& Nguyen (2018) and Basri (2018), either single- or double-dip signatures are more likely an effect of hemispheric asymmetries caused by the presence of a few or several spots, their surface distribution and dynamics. Rotational modulation also often presents long-term amplitude variations associated to activity cycles (e.g.: Ferreira Lopes et al. 2015) that use to be somewhat irregular, as well as showing some asymmetry with respect to the flux average. On the other hand, pulsation typically displays a more regular shape of the flux variation and may have constant amplitude, or a regular amplitude variation usually forming steady beats (which can be clearly seen, in particular, in the wavelet maps). Some pulsators, such as for example Gamma Douradus variables, may present some irregularities in their LCs that can be confused with rotational modulation. Those cases can be disentangled when presenting an asymmetry in their variability signatures skewed to higher fluxes, differently of the rotational modulation that tends to present an asymmetry skewed to lower fluxes. A few cases are difficult to unravel, so an additional analysis considering the stellar physical parameters is performed to identify their natures. When no conclusion can be taken for some cases, then an ambiguous variation is set to their LC classifications.

\subsection{The Fast Fourier Transform Analysis }

The Fast Fourier Transform (FFT), a computationally faster version of the Discrete Fourier Transform (DFT), is a discrete version of the continuous Fourier Transform that can decompose periodicities of real data. The algorithm used in this work is based on Cooley \& Tukey (1965) and similar to the one employed by Sanchis-Ojeda et al. (2013, 2014), but is optimized for the TESS data. The main advantage of FFT is its computational speed, which is not a requirement for our purposes because our sample is not too large. We simply use FFT as a complementary method for interpreting periodicities. As a limitation, FFT  is applicable for evenly-sampled time series. The TESS LCs are nearly evenly sampled with a few irregularities, especially when having some gaps. To ensure that the LCs are evenly spaced, we rebinned the data to regular time intervals close to their original bins and fulfilled eventual gaps with linear interpolation.

\subsection{The Lomb-Scargle Analysis}

Lomb-Scargle (Lomb 1976; Scargle 1982) is a well-known algorithm that can provide Fourier-like periodograms of real (discrete) data. Its main advantage over FFT is the fact that it can deal with unevenly-sampled time series. Periodograms can thus be obtained directly from the LCs with their original time samplings, without the need of any rebinning or interpolation. Another useful feature, developed by Horne \& Baliunas (1986), is a formal calculation, inherent to the method, of false alarm probabilities (FAP) for detected periods. Such a statistics helps us in quantitatively validating periods. Stellar variability periods were then identified by the main periodogram peaks with confidence levels greater than 99\% (De Medeiros et al. 2013).

\subsection{The Wavelet Analysis} \label{wavelet_analysis}

The wavelet transform (e.g., Grossmann \& Morlet 1984) is a powerful tool to analyze a time series in the time-frequency domain, namely by decomposing periodicities as power spectra sections along the time window of the data. This method is comparable to the short-term Fourier transform (STFT) (Gabor 1946), which decomposes a time series into Fourier transforms of short-term boxes along the time window. The boxes in the STFT, however, have fixed lengths that may typically hinder lower-frequency signals if a high temporal resolution is aimed. The wavelet transform overcomes such a resolution issue by convolving the time series with an orthonormal function called {\em mother wavelet} with variable dilation and translation parameters that self-adjust to the different frequencies of a signal. The time-frequency diagram of a wavelet transform, namely the wavelet map or local wavelet spectrum, thus decomposes a signal into all frequencies naturally, within a region of confidence, without the need of defining some box length. In addition, a global power spectrum can also be obtained by integrating the wavelet map along the time axis. This global wavelet spectrum gives us a view of the main periodicities present in a time series that can be compared with other power spectra, such as those from FFT and Lomb-Scargle. Overall, the wavelet technique is a useful tool for analyzing non-stationary and non-periodic signals, revealing characteristics that can vary in both time and frequency (Burrus, Gopinath \& Guo 1998). To date, a plethora of problems in Astronomy, mostly associated to the search for periodicities have been treated on the basis of the wavelet technique (e.g.: Espaillat et al. 2008; Bravo et al. 2014; Mathur et al. 2014; Bewketu-Belete, A. et al. 2018; de Lira et al. 2019; Santos et al. 2019; Reinhold \& Hekker 2020).

The wavelet maps can reveal detailed signatures of a variability behavior that may not be evident in the time series itself or in global power spectra. Therefore, the wavelet method helps us very much in the identification of the types of variability identified in a LC. We refer to Bravo et al. (2014) for a detailed analysis of different signatures that can be observed in wavelet maps of stellar LCs. An important example is the case of analyzing double-dip rotational modulations (Basri \& Nguyen 2018) to obtain proper rotation periods rather than aliases. The typical signature of such a case observed in the wavelet map is the presence of two dominant features along time, the period of a feature being the double or a half of the other. In many cases, the rotation period tends to be the longer-period feature, the shorter-period one being an effect of the superposition of two semi-sinusoids associated to the double-dip signature. Figure 5 from Bravo et al. (2014) illustrates a typical example of such a case. Nevertheless, careful inspection of the LCs along with different tools is necessary for a proper conclusion of the actual period.

\subsection{Visual inspection} 

\begin{figure}[h!]
	\centering
	\includegraphics[scale=.8]{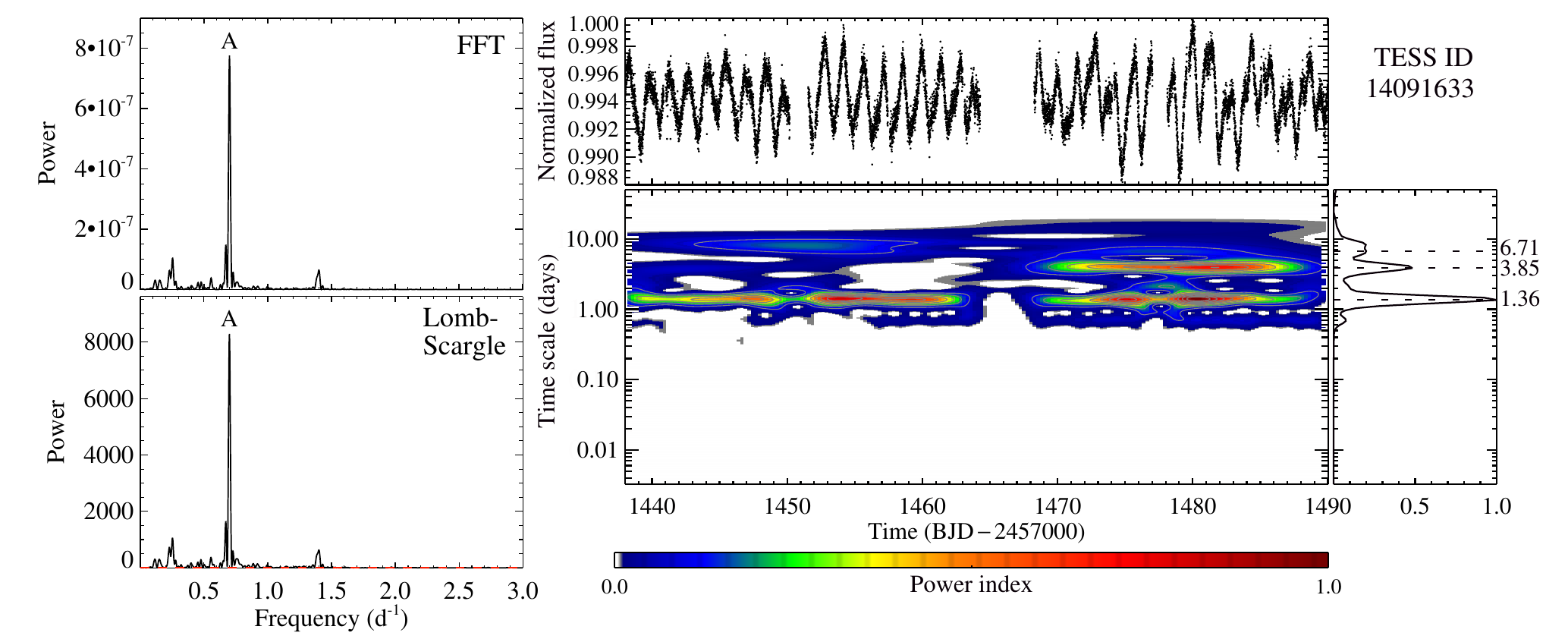}
	\includegraphics[scale=.8]{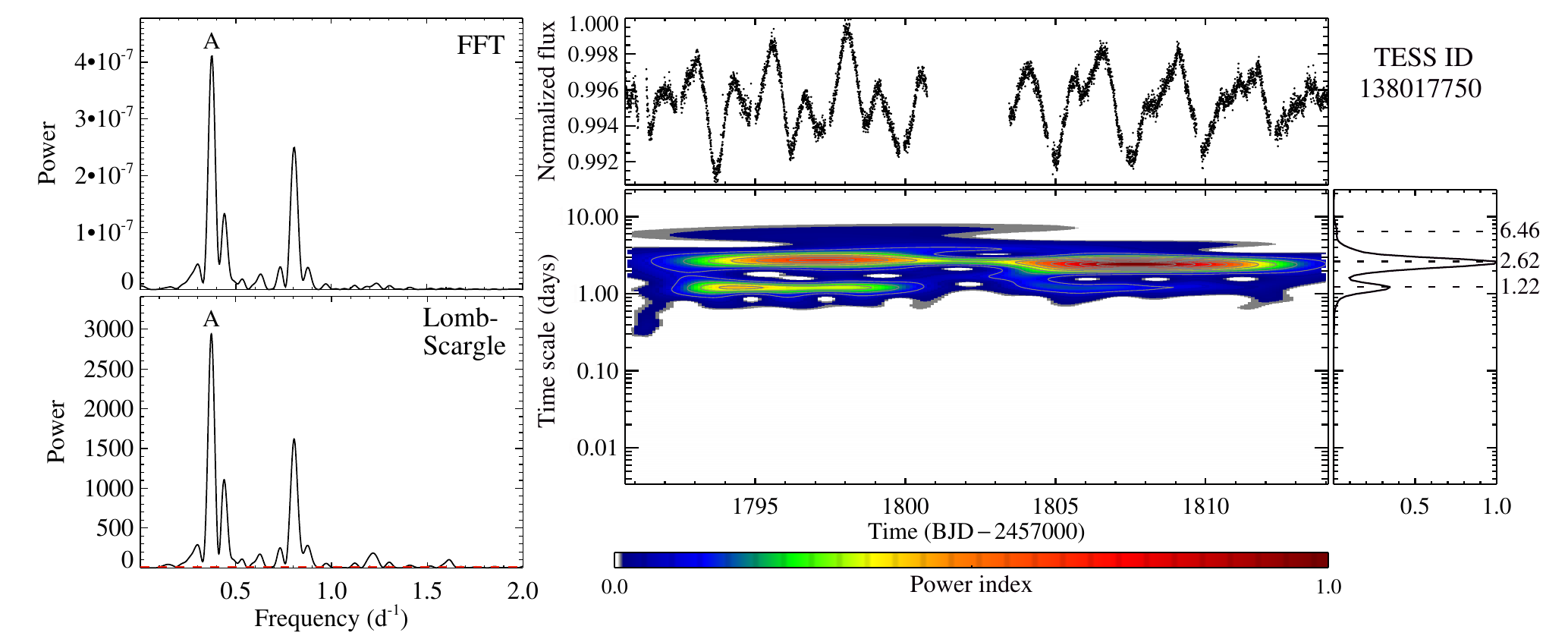}
	\includegraphics[scale=.8]{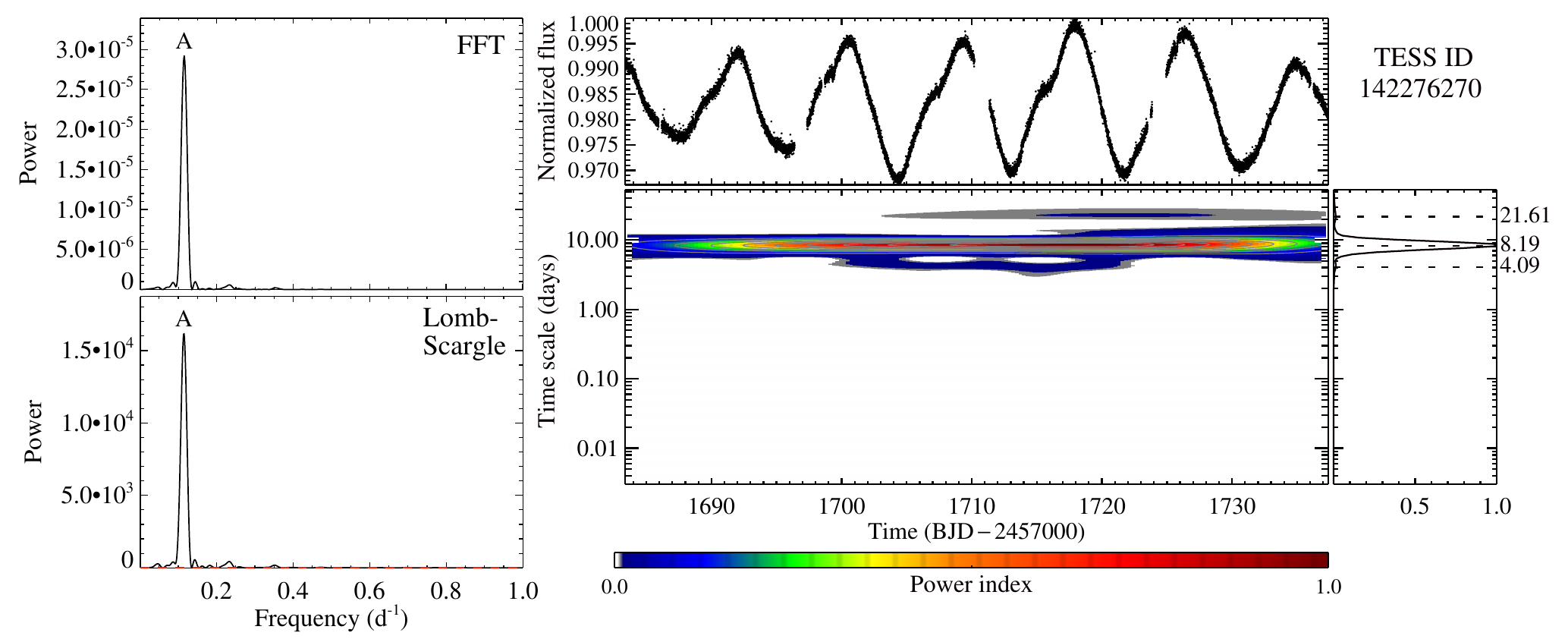}
	\caption{Examples of diagnostic plots displaying FFT and Lomb-Scargle periodograms, LCs and wavelet maps for three TOIs with typical rotation signatures. Persistent periods of 1.361, 2.623, and 8.189 days, respectively, for TIC 14091633 (top panels), TIC 138017750 (middle panels), and TIC 142276270 (bottom panels), are observed in their wavelet maps and confirmed by FFT and Lomb-Scargle peaks labeled {\em A}. The complete figure set (131 images) is available in the online journal.} 
	\label{rot}
\end{figure}

Once the FFT, Lomb-Scargle and wavelet results in hands, we perform a visual inspection on each LC to identify effective modulation traces based on the procedure applied by De Medeiros et al. (2013). Readers are referred to Sect. 2.1 for a short description of the signatures searched in this work and to Sect. 2.2.2 of De Medeiros et al. (2013) for a detailed discussion on such procedure. Following those authors, we considered that stars with more than three observed cycles in their LCs have confident periods, where the effective number of cycles (N$_{Cycle}$) is the effective time span (t$_{SPAN}$) of the LC, excluding gaps, divided by the rotation period (P$_{rot}$). Neverthless, stars with 2.5 $\leq$ N$_{Cycle} <$ 3.0 whose LCs show clear rotation signature with large-amplitude fluctuation, persistent all along the effective time span, were also considered to have confident periods. Figure \ref{rot} displays examples of LCs presenting typical rotation signature identified in our sample, with the corresponding FFT and Lomb-Scargle periodograms, as well as the wavelet maps. Figures following the same design of Fig. \ref{rot} are provided in the online material for all the stars with rotation and other variability signatures revealed by our analysis.

\figsetstart
\figsetnum{1}
\figsettitle{Diagnostic plots for the 131 TOIs with unambiguous rotation periods}

\figsetgrpstart
\figsetgrpnum{1.1}
\figsetgrptitle{TIC 2760710}
\figsetplot{2760710.pdf}
\figsetgrpnote{Persistent period of 1.251 day for TIC 2760710 is observed in the wavelet map and confirmed by FFT and Lomb-Scargle peaks labeled A.}
\figsetgrpend

\figsetgrpstart
\figsetgrpnum{1.2}
\figsetgrptitle{TIC 7624182}
\figsetplot{7624182.pdf}
\figsetgrpnote{Persistent period of 1.624 day for TIC 7624182 is observed in the wavelet map and confirmed by FFT and Lomb-Scargle peaks labeled A.}
\figsetgrpend

\figsetgrpstart
\figsetgrpnum{1.3}
\figsetgrptitle{TIC 9033144}
\figsetplot{9033144.pdf}
\figsetgrpnote{Persistent period of 4.201 days for TIC 9033144 is observed in the wavelet map and confirmed by FFT and Lomb-Scargle peaks labeled A.}
\figsetgrpend

\figsetgrpstart
\figsetgrpnum{1.4}
\figsetgrptitle{TIC 9348006}
\figsetplot{9348006.pdf}
\figsetgrpnote{Persistent period of 5.329 days for TIC 9348006 is observed in the wavelet map and confirmed by FFT and Lomb-Scargle peaks labeled A.}
\figsetgrpend

\figsetgrpstart
\figsetgrpnum{1.5}
\figsetgrptitle{TIC 13499636}
\figsetplot{13499636.pdf}
\figsetgrpnote{Persistent period of 5.595 days for TIC 13499636 is observed in the wavelet map and confirmed by FFT and Lomb-Scargle peaks labeled A.}
\figsetgrpend

\figsetgrpstart
\figsetgrpnum{1.6}
\figsetgrptitle{TIC 14091633}
\figsetplot{14091633.pdf}
\figsetgrpnote{Persistent period of 1.363 day for TIC 14091633 is observed in the wavelet map and confirmed by FFT and Lomb-Scargle peaks labeled A.}
\figsetgrpend

\figsetgrpstart
\figsetgrpnum{1.34}
\figsetgrptitle{TIC 138017750}
\figsetplot{138017750.pdf}
\figsetgrpnote{Persistent period of 2.623 days for TIC 138017750 is observed in the wavelet map and confirmed by FFT and Lomb-Scargle peaks labeled A.}
\figsetgrpend

\figsetgrpstart
\figsetgrpnum{1.35}
\figsetgrptitle{TIC 138588540}
\figsetplot{138588540.pdf}
\figsetgrpnote{Persistent period of 9.236 days for TIC 138588540 is observed in the wavelet map and confirmed by FFT and Lomb-Scargle peaks labeled A.}
\figsetgrpend

\figsetgrpstart
\figsetgrpnum{1.36}
\figsetgrptitle{TIC 140068425}
\figsetplot{140068425.pdf}
\figsetgrpnote{Persistent period of 1.064 day for TIC 140068425 is observed in the wavelet map and confirmed by FFT and Lomb-Scargle peaks labeled A.}
\figsetgrpend

\figsetgrpstart
\figsetgrpnum{1.37}
\figsetgrptitle{TIC 142090065}
\figsetplot{142090065.pdf}
\figsetgrpnote{Persistent period of 2.376 days for TIC 142090065 is observed in the wavelet map and confirmed by FFT and Lomb-Scargle peaks labeled A.}
\figsetgrpend

\figsetgrpstart
\figsetgrpnum{1.38}
\figsetgrptitle{TIC 142276270}
\figsetplot{142276270.pdf}
\figsetgrpnote{Persistent period of 8.189 days for TIC 142276270 is observed in the wavelet map and confirmed by FFT and Lomb-Scargle peaks labeled A.}
\figsetgrpend

\figsetend

\section{Results} \label{sec:results}

\begin{figure}[h!]
	\centering
	\includegraphics[scale=.8]{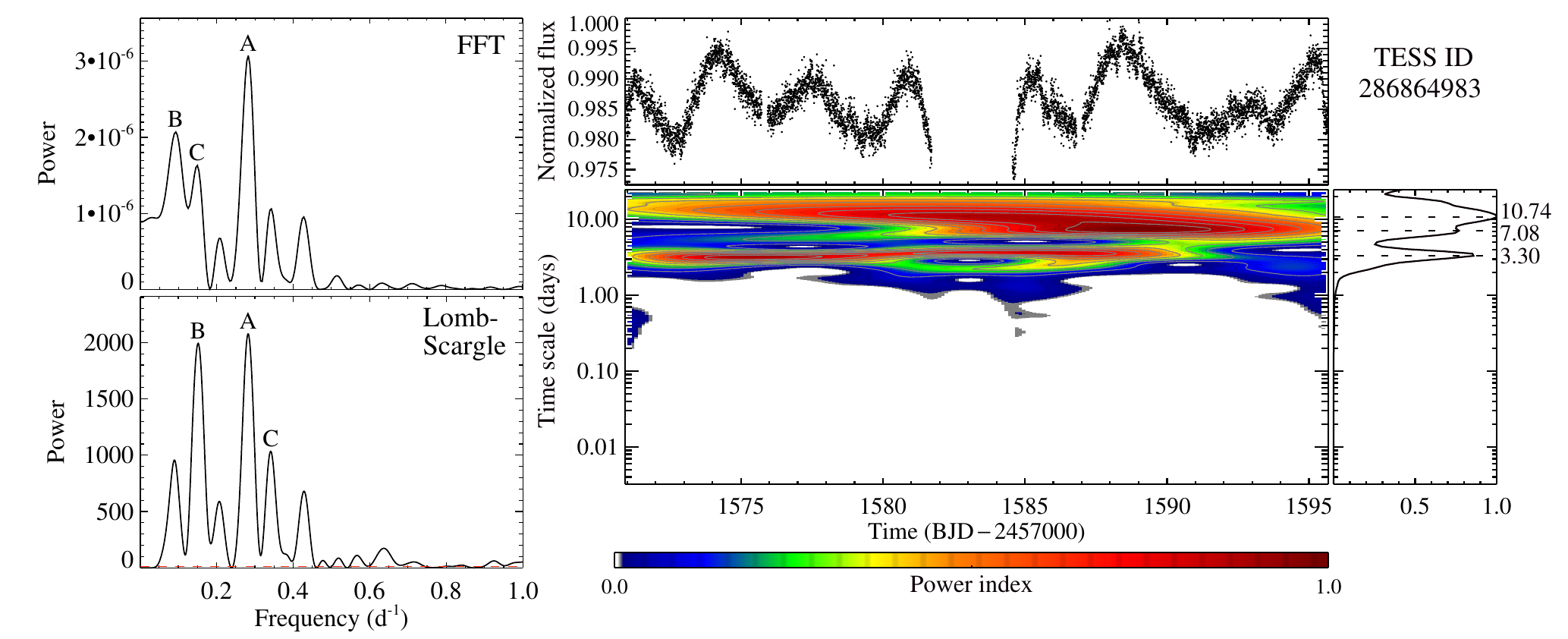}
	\includegraphics[scale=.8]{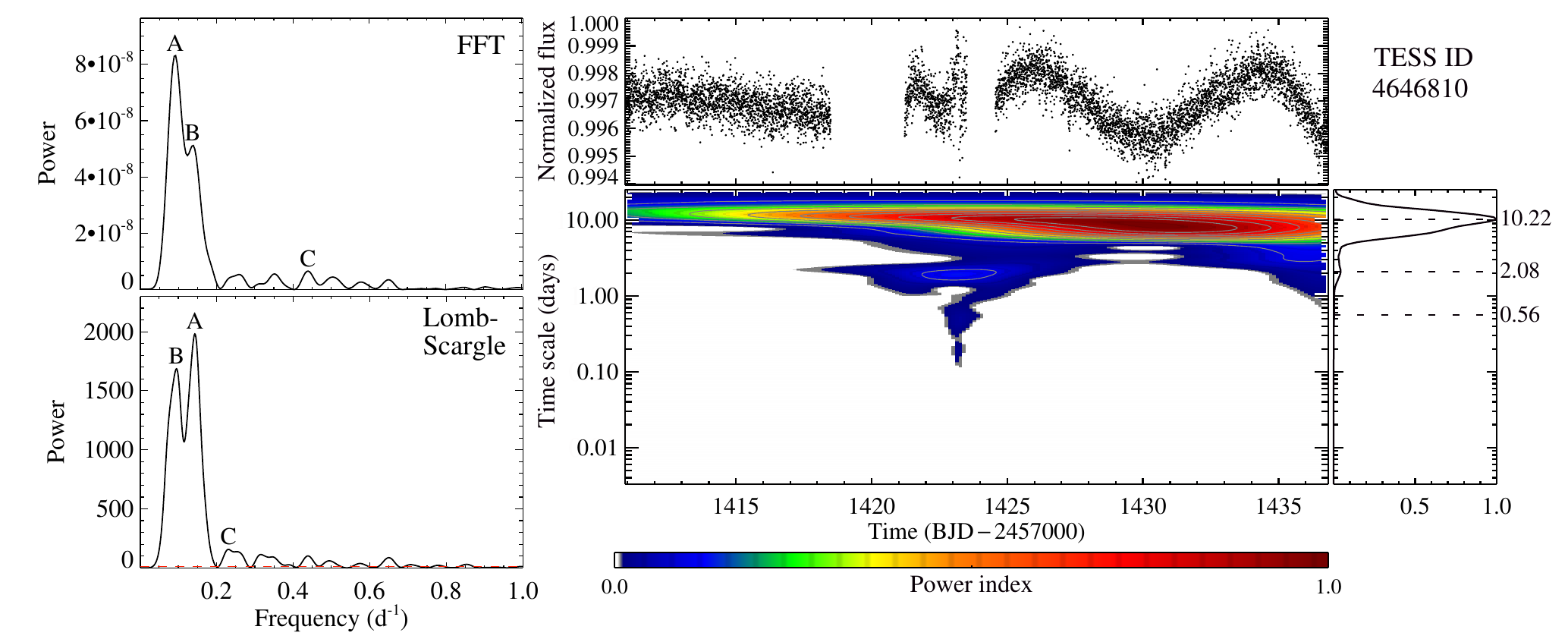}
	\caption{Examples of diagnostic plots displaying FFT and Lomb-Scargle periodograms, LCs and wavelet maps for two TOIs showing typical characteristics of dubious rotation periods (top panels, TIC 286864983) and ambiguous variability (bottom panels, TIC 4646810). In the top panels, in spite of a potential rotation signature, periodograms and wavelet map reveal a multiple periodicity of no clear diagnosis. In the bottom panels, the LC seems irregular with no apparent variability at the beginning, some systematics in the middle, and a possible variability at the second half. The complete figure set (141 images) is available in the online journal.
	} 
	\label{dub_var}
\end{figure}

We have analyzed a total of 1000 targets presenting public LCs, with short-cadence TESS observations in sectors 1 to 22, classified as TESS Objects of Interest. Among those stars, we have identified 163 targets with rotation signature, including 131 with unambiguous rotation periodicities, 32 targets with rotation signature but having dubious values for the periods, and  109 stars with ambiguous variability. Dubious rotation periods correspond to stars showing potential rotation signature, but whose period could not be disentangled among two or more possibilities (from periodogram peaks and wavelet maps), as well as stars with N$_{Cycle} <$ 3, except for some cases with 2.5 $<$ N$_{Cycle} <$ 3 that show clear and persistent rotation pattern along their LCs (see Sect. 2.5). Ambiguous variability corresponds to stars showing visually noticeable fluctuations that are faint for proper interpretation or with insufficient time span for proper signature identification, as well as significant large-amplitude variations with a very irregular or complex behavior usually caused by systematics. Some clear variabilities may eventually be classified as ambiguous when could not be discriminated among rotation, pulsation or other signature, as described in Sect. 2.1, and those cases shall be revisited in future works, especially using additional observations. Figure \ref{dub_var} displays typical examples of LCs with dubious rotation periods and ambiguous variability. Table \ref{tab_unamb_rot} lists stars with unambiguous rotation periodicities. For each star, from left to right, the columns show the following: the TIC ID, stellar coordinates, stellar parameters (T$_{eff}$ and $\log g$), orbital period (P$_{orb}$), rotation period (P$_{rot}$), error in the rotation period ({\em e}P$_{rot}$), effective time span ($t_{SPAN}$) of each LC (the total time span subtracted by the duration of eventual gaps), the effective number of cycles of the rotational modulation (defined as N$_{Cycle}$ = $t_{SPAN}$/$P_{rot}$), and the TESS observation sectors. Table \ref{dubious} lists the TOIs with dubious rotation periods, whereas Table \ref{ambiguous} lists the stars with ambiguous variability. 

\figsetstart
\figsetnum{2}
\figsettitle{Diagnostic plots for the TOIs with dubious rotation periods and ambiguous variability}

\figsetgrpstart
\figsetgrpnum{2.23}
\figsetgrptitle{TIC 286864983}
\figsetplot{286864983.pdf}
\figsetgrpnote{TOI with dubious rotation period TIC 286864983. Possible period of 3.30 or 7.08 days.}
\figsetgrpend

\figsetgrpstart
\figsetgrpnum{2.35}
\figsetgrptitle{TIC 4646810}
\figsetplot{4646810.pdf}
\figsetgrpnote{TOI with ambiguous variability TIC 4646810.}
\figsetgrpend

\figsetend


\begin{deluxetable*}{cccccccccc}
\tablenum{1}
\tablecaption{Catalog of TOIs with unambiguous rotation periods from our analysis.\label{tab_unamb_rot}}
\tablewidth{2pt}
\tablehead{
\colhead{TIC ID} & \colhead{T$_{eff}$} & \colhead{$\log g$	} & \colhead{P$_{orb}$} &
\colhead{P$_{rot}$} & \colhead{{\em e}P$_{rot}$} & \colhead{t$_{SPAN}$} & \colhead{N$_{Cycle}$} & \colhead{Sectors} \\
\colhead{}	&	\colhead{(K)}	&	\colhead{(cm/s$^2$)}	&	\colhead{(days)}	&	\colhead{(days)}	&	\colhead{(days)}	&	\colhead{(days)}	&	\colhead{}&	\colhead{}
}
\startdata
2760710	&	2808	&	5.206	&		&	1.251	&	0.033	&	24	&	19.2	&	 2 \\
7624182	&	8666	&	3.801	&	1.108	&	1.624	&	0.029	&	45	&	27.7	&	 4 \\
9033144	&	5757	&	3.900	&	4.715	&	4.201	&	0.368	&	24	&	5.7	&	 2 \\
9348006	&	5251	&	4.543	&	10.240	&	5.329	&	0.592	&	24	&	4.5	&	 21 \\
13499636	&	5518	&	4.592	&	11.325	&	5.595	&	0.921	&	17	&	3.0	&	 15 \\
14091633	&	6350	&	4.340	&	5.529	&	1.363	&	0.022	&	43	&	31.5	&	 5 \\
...	&	...	&	...	&	...	&	...	&	...	&...	&	...	&	...	\\
\enddata
\tablecomments{With one row for each TOI, the following information is listed: the TIC ID, effective temperature (T$_{eff}$), surface gravity ($\log g$), and orbital period (P$_{orb}$) taken from the TOI Release Portal (\url{https://tess.mit.edu/toi-releases/}), rotation period (P$_{rot}$), error in the rotation period ({\em e}P$_{rot}$), effective time span (t$_{SPAN}$), and effective number of cycles (N$_{Cycle}$), obtained from our analysis, and TESS observation sectors. Values for $\log g$ and P$_{orb}$ are rounded to 3 decimals digits. The complete table is provided in machine-readable form in the online journal. Here we show a fragment for guidance regarding its form and content. }
\end{deluxetable*}

\begin{deluxetable*}{cccccccc}
\tablenum{2}
\tablecaption{List of the 32 TOIs with dubious rotation periods from our analysis.\label{dubious}}
\tablewidth{2pt}
\tablehead{
\colhead{TIC ID} & \colhead{T$_{eff}$} & \colhead{$\log g$	} & \colhead{P$_{orb}$} &
\colhead{P$_{rot}$} & \colhead{t$_{SPAN}$} & \colhead{N$_{Cycle}$} & \colhead{Sectors} \\
\colhead{}	&	\colhead{(K)}	&	\colhead{(cm/s$^2$)}	&	\colhead{(days)}	&	\colhead{(days)}	&	\colhead{(days)}		&	\colhead{(days)}		&\colhead{}
}
\startdata
1129033	&	5500	&	4.483	&	1.360	&	 5.00/10.00 	&	19	&	1.9	&	 4 \\
1528696	&	4975	&	4.520	&	0.882	&	 5.15/8.97 	&	23	&	2.6	&	 5 \\       
9006668	&	5024	&	4.569	&	1.272	&	 7.05/9.96 	&	25	&	2.5	&	 2 \\       
35516889	&	5568	&	4.393	&	0.789	&	 6.18/9.37 	&	19	&	2.0	&	 9 \\       
36734222	&	4400	&	4.646	&	0.813	&	 7.41$^a$ 	&	19	&	2.6	&	 9 \\
62483237	&	4356	&	4.535	&	11.058	&	 6.83/11.09 	&	24	&	2.2	&	 1 \\
... & ... & ...		&	...	&	...	&		...		&		...		&	...	\\
\enddata
\tablecomments{The following information is listed: the TIC ID, effective temperature (T$_{eff}$), surface gravity ($\log g$), and orbital period (P$_{orb}$) taken from the TOI Release Portal (\url{https://tess.mit.edu/toi-releases/}), likely rotation period values (P$_{rot}$), effective time span (t$_{SPAN}$), and effective number of cycles (N$_{Cycle}$), obtained from our analysis, and TESS observation sectors. Flag {\em a} corresponds to stars with less than three observed cycles that show non-persistent pattern along their LCs. Values for $\log g$ and P$_{orb}$ are rounded to 3 decimals digits. The complete table is provided in machine-readable form in the online journal. Here we show a fragment for guidance regarding its form and content.}
\end{deluxetable*}

\begin{deluxetable*}{ccccccc}
\tablenum{3}
\tablecaption{List of the 109 TOIs with ambiguous variability behavior from our analysis.\label{ambiguous}}
\tablewidth{2pt}
\tablehead{
\colhead{TIC ID} & \colhead{T$_{eff}$} & \colhead{$\log g$	} & \colhead{P$_{orb}$} & \colhead{t$_{SPAN}$} & \colhead{Sectors} \\
\colhead{}	&	\colhead{(K)}	&	\colhead{(cm/s$^2$)}	&	\colhead{(days)}	&	\colhead{(days)}	&	\colhead{}
}
\startdata
1003831	&	5752	&	4.471	&	1.651	&	18	&	 8 \\       
1103432	&	6231	&	4.264	&	3.728	&	17	&	 8 \\       
4646810	&	4884	&	4.490	&	14.490	&	21	&	 4 \\       
9804616	&	3274	&	4.979	&	0.517	&	19	&	 4 \\
12862099	&	5410	&	4.479	&	2.424	&	17	&	 3 \\       
13684720	&	3275	&	4.758	&	12.438	&	36	&	 14,15 \\       
...& ...& ...	&	...	&	...	&	...	\\							
\enddata
\tablecomments{The following information is listed: the TIC ID, effective temperature (T$_{eff}$), surface gravity ($\log g$), and orbital period (P$_{orb}$) taken from the TOI Release Portal (\url{https://tess.mit.edu/toi-releases/}), effective time span (t$_{SPAN}$) obtained from our analysis, and TESS observation sectors. Values for $\log g$ and P$_{orb}$ are rounded to 3 decimals digits. The complete table is provided in machine-readable form in the online journal. Here we show a fragment for guidance regarding its form and content.}
\end{deluxetable*}

The fraction of stars that show rotational modulation is 16\% of the parent sample of 1000 TOIs considered in this work. Indeed, the detection of stellar variability it depends strongly on instrumental characteristics, such as photometric sensitivity, time span of the observation (see, e.g., Le\~ao et al. 2015), and even on LCs reduction and treatment procedures used (de Lira et al. 2019). For instance, rotational modulation was detected for no more than 5\% of the total sample of CoRoT stars (e.g., Meibom et al. 2011; De Medeiros et al. 2013), whereas for the total sample of {\em Kepler} stars that fraction increased to about 20\% (e.g., Nielsen et al. 2013; McQuillan et al. 2013a,b, 2014; Reinhold et al. 2013; Walkowicz \& Basri 2013; Paz-Chinch\'on et al. 2015; Reinhold \& Hekker 2020). The rotation signature detected in 16\% of the stars in our sample is in agreement with that found in {\em Kepler} stars. Among those targets with unambiguous rotation, the following targets exhibit potential flare events with the date of the major feature indicated: TIC 200322593 (Nov 25, 2018), TIC 233211762 (Nov 9, 2019), TIC 244161191 (Oct 3, 2018), TIC 278198753 (May 27, 2019), TIC 300293197 (Nov 21, 2018), TIC 307610438 (May 1, 2019), TIC 318937509 (Jan 16, 2019), TIC 32830028 (Dec 23, 2018), TIC 348538431 (Jan 28, 2019), TIC 460205581 (May 3, 2019), TIC 47384844 (Mar 11, 2019), TIC 67646988 (Feb 12, 2020), TIC 77951245 (Nov 19, 2018), and TIC 93125144 (Dec 22, 2018). Six other stars with unambiguous rotation, TIC 70797900, TIC 235037761, TIC 299798795, TIC 206609630, and TIC 410214986 also exhibit flare events as previously reported by G\"unther et al. (2020), as well as TIC 98796344  and TIC 257605131, reported by Howard et al. (2019) and Tu et al. (2020), respectively. In addition, one star with dubious rotation period, TIC 233120979, and two stars with ambiguous variability, TIC 13684720 and TIC 89256802, show flare events with the major one at September 9th 2019, August 2nd 2019, and November 10th 2018, respectively. G\"unther et al. (2020) also reported flare events for TIC 32090583.

It is worthy to underline the large number of 714 TOIs exhibiting a noisy behavior in their LCs, corresponding to 71\% of the parent sample. Although those stars present typically low-amplitude signals whose physical periodicities cannot be easily identified, from a certain view they can also point for key information. Typically, a noisy signature is a complex combination of instrumental noise contributions (related, for instance, with Poisson statistics and readout noise) plus a relevant contribution of intrinsic stellar noise, Galactic position, light from neighboring stars and sky background contamination (e.g.: Gilliland et al. 2011). When the TOI LCs considered in this work present a low-amplitude signal, we assume them to be a noisy signature. Nevertheless, for some stars the noisy behavior could reflect low activity or long periodicities, in particular for those targets with short observational time span. It should be noticed that part of the stars classified as having noisy LCs may have been set up this way because of data reduction issues. As such, caution should be taken with this subsample when using it for planet search strategies. Additional observations and data treatments may change the status of some of these stars. Table \ref{noisy} lists the stars with noisy LCs. Among those targets with noisy LCs, six of them, TIC 186812530, TIC 230086768, TIC 286865921, TIC 365639282, and TIC 36622912 exhibit potential flare events, with major features at January 28th 2019, September 2nd 2019, April 17th 2019, December 16th 2018, and January 16th 2019, respectively. The noisy-LC star TIC 272086159 also exhibits flare events as reported by G\"unther et al. (2020).

\begin{deluxetable*}{cccccc}
\tablenum{4}
\tablecaption{List of the 714 TOIs with noisy LCs from our analysis.\label{noisy}}
\tablewidth{2pt}
\tablehead{
\colhead{TIC ID} & \colhead{T$_{eff}$} & \colhead{$\log g$	} & \colhead{P$_{orb}$} & \colhead{t$_{SPAN}$} & \colhead{Sectors} \\
\colhead{}	&	\colhead{(K)}	&	\colhead{(cm/s$^2$)}	&	\colhead{(days)}	&	\colhead{(days)}	&	\colhead{}
}
\startdata
1133072	&	3380	&	4.925	&	0.847	&	15	&	 8	\\
1449640	&	6383	&	4.030	&	3.502	&	23	&	 5	\\
4616072	&	6675	&	4.201	&	4.186	&	40	&	 6	\\
4897275	&	5854	&	4.386	&	16.710	&	24	&	 21	\\
5868998	&	3602	&	4.817	&	0.636	&	18	&	 10	\\
6663331	&	5498	&	4.479	&	3.180	&	23	&	 13	\\
...		&		...		&		...	&	...	&	...	&	...	\\
\enddata
\tablecomments{The following information is listed: the TIC ID, effective temperature (T$_{eff}$), surface gravity ($\log g$), and orbital period (P$_{orb}$) taken from the TOI Release Portal (\url{https://tess.mit.edu/toi-releases/}), effective time span (t$_{SPAN}$) obtained from our analysis, and TESS observation sectors. Values for $\log g$ and P$_{orb}$ are rounded to 3 decimals digits. The complete table is provided in machine-readable form in the online journal. Here we show a fragment for guidance regarding its form and content.}
\end{deluxetable*}

\begin{figure}[h!]
	\centering
	\includegraphics[scale=.8]{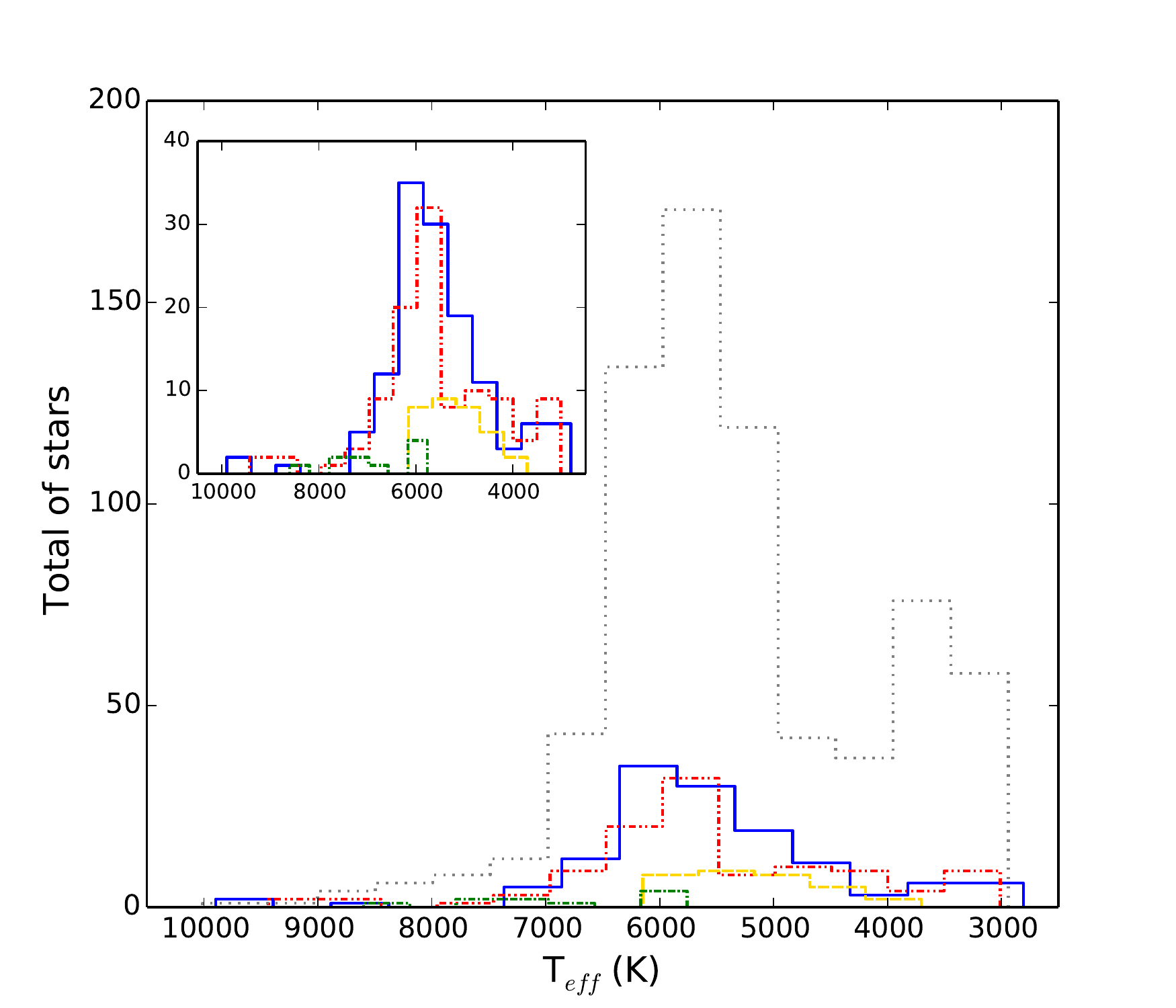}
	\caption{The distribution of the effective temperature for the TOIs analyzed in the present study. Lines in solid blue, dashed orange, dot-dashed green, and dot-dot-dashed red, and dashed gray are for stars with unambiguous rotation periods, dubious rotation periods, pulsation, and ambiguous variability, and noisy LCs, respectively. A closer view of the distribution of the effective temperature without the subsample of stars with noisy LCs is displayed in the upper left corner of the figure for better visualization.} 
	\label{Teff}
\end{figure}

\begin{figure}[h!]
	\centering
	\includegraphics[scale=.8]{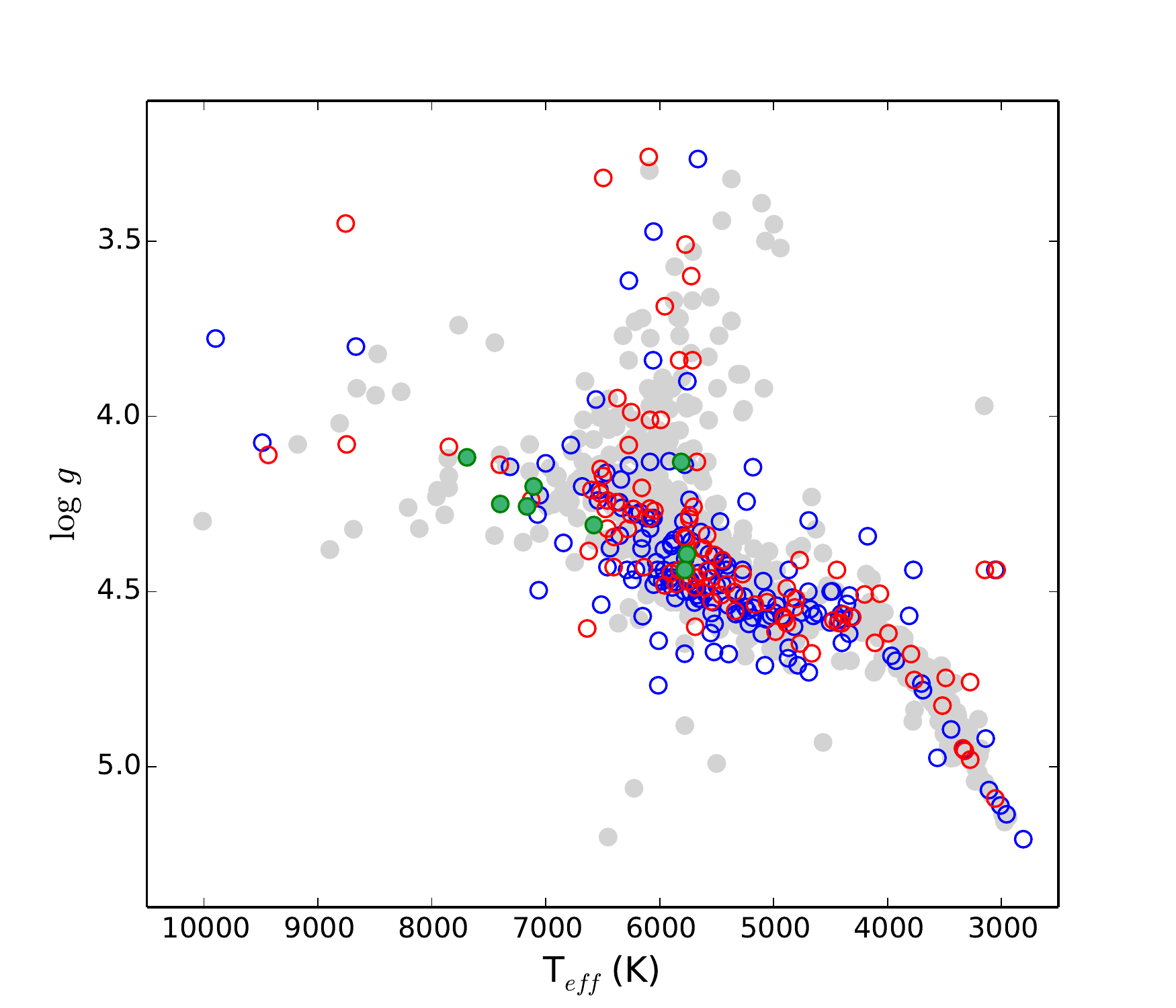}
	\caption{The T$_{eff}$ versus $\log g$ diagram for the sample of 1000 TOIs composing the present study. Circles in open blue, solid green, open red, and solid gray are for stars with unambiguous and dubious rotation periods, pulsation, ambiguous variability, and noisy LCs, respectively. The distributions of TOIs with rotation, ambiguous variability, and noisy LCs follow fairly the same scenario.} 
	\label{HR}
\end{figure}

\begin{figure}[h!]
	\centering
	\includegraphics[scale=.8]{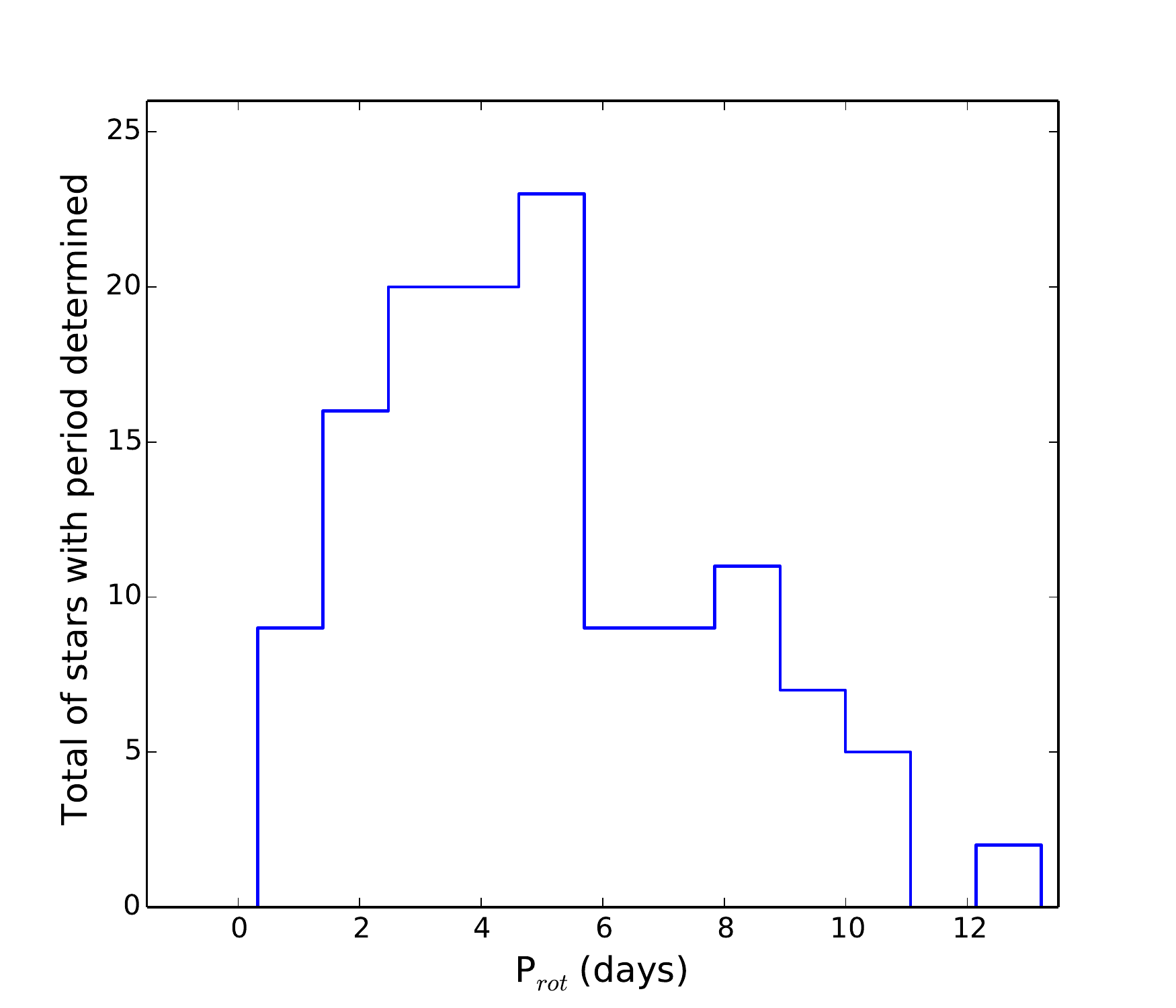}
	\caption{The rotation period distribution for the 131 TOI stars exhibiting unambiguous rotation signatures from the present analysis.} 
	\label{Prot}
\end{figure}

Based on the rotation periods and other stellar parameters listed in Tabs. \ref{tab_unamb_rot} to \ref{noisy}, the following major scenarios emerge. First, the whole sample of 1000 TOIs covers a range of effective temperature from 2,808 K to 9,898 K, typically stars of spectral type from M6 to A0, a scenario followed by the stars with rotation signature, ambiguous variability and noisy LCs. Figure \ref{Teff} illustrates the effective temperature distributions for the underlined samples. Second, the distribution of the different subsample of stars, namely stars with rotation signature (with unambiguous and dubious periodicities), stars showing ambiguous stellar variability, and stars with noisy behavior, follow, approximately, the same trend in the $\log g$ versus T$_{eff}$ diagram as displayed in Fig. \ref{HR}. Third, as it arises in Fig. \ref{Prot}, the distribution of the rotational periods ranges between 0.321 and 13.219 days. Overall, the range of this distribution is associated to the TESS technical limits of 28 days baseline per sector, which does not favor the determination of longer periods of rotation, also common among M dwarf stars (e.g.: Newton et al. 2018; Oelkers et al. 2018), but only periods shorter than 28 days. Even for the LCs obtained from combined sectors, thus with long time spans, the post-treatments needed in this process may hinder longer periodicities. The rotation period distribution also reveals a trend for a bimodality, with a peak around 5 days and a second one arising around 8 days. Such a trend reflects what is expected for cool stars, as reported by McQuillan (2013, 2014) and Davenport (2017). However, caution should be taken in its interpretation, which could be associated to the present sample limitation, especially at lower temperatures.

\begin{figure}[h!]
	\centering
	\includegraphics[scale=.8]{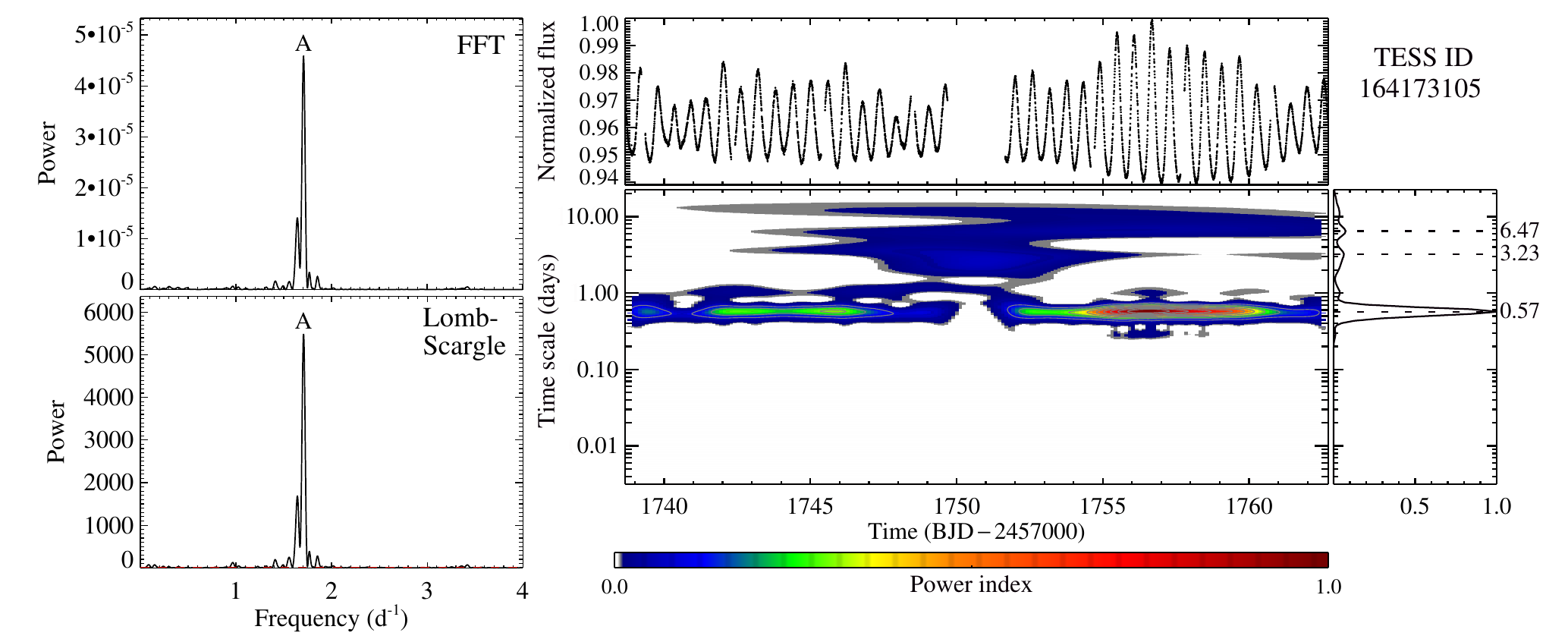}
	\includegraphics[scale=.8]{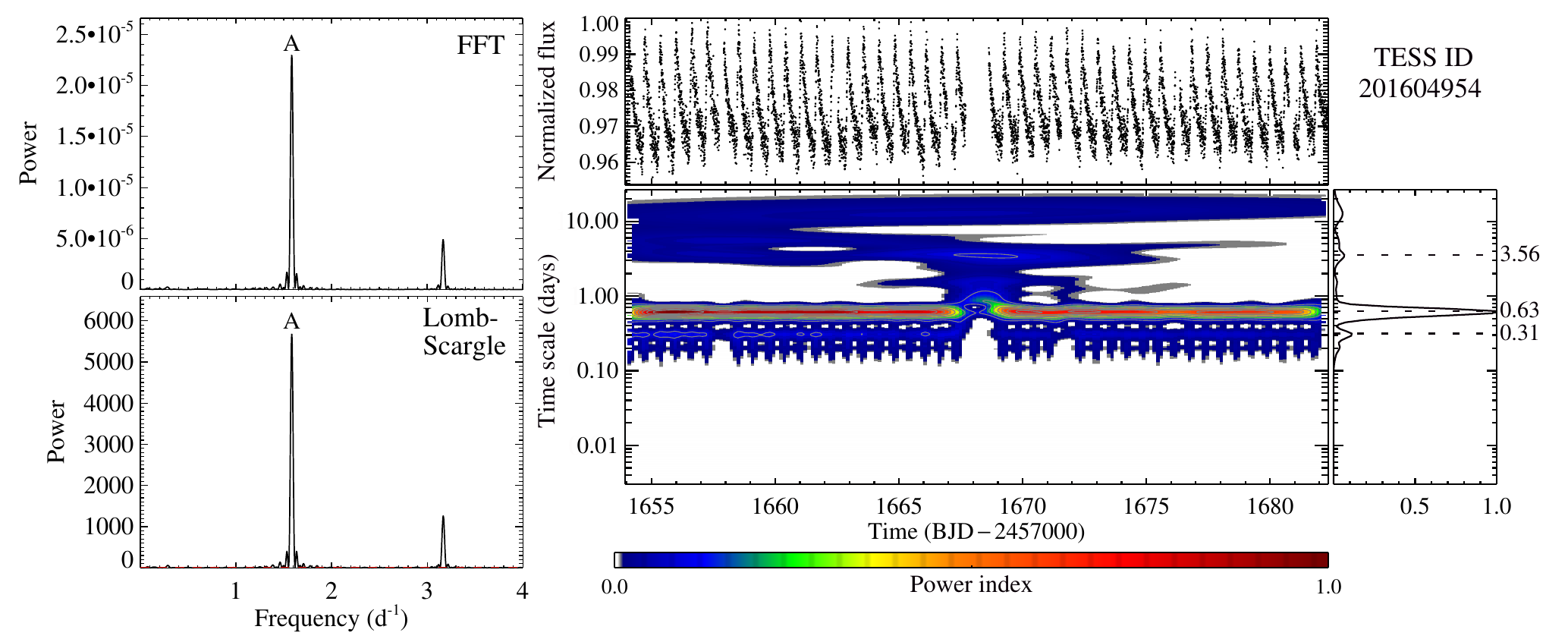}
	\includegraphics[scale=.8]{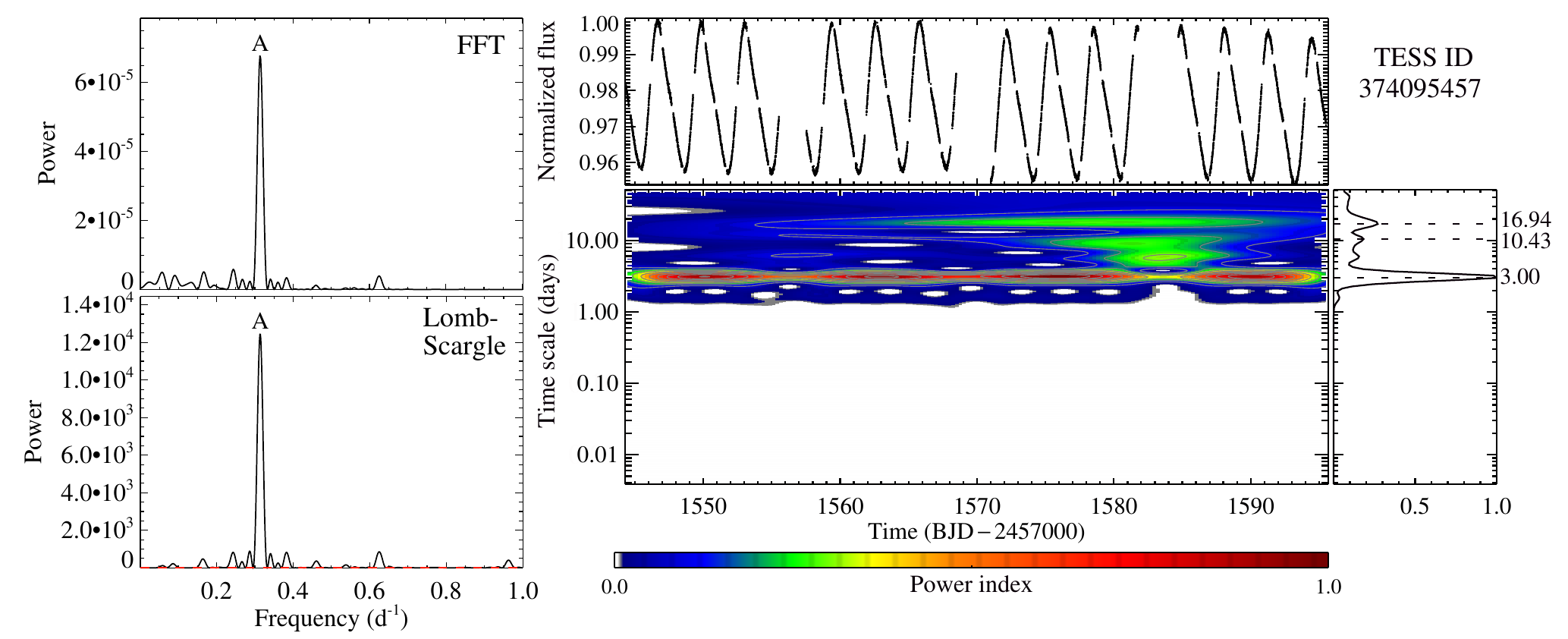}
	\caption{Examples of diagnostic plots displaying FFT and Lomb-Scargle periodograms, LCs and wavelet maps for three TOIs with typical pulsation signatures. Persistent periods of 0.572, 0.629, and 2.995 days, respectively, for TIC 164173105 (top panels), TIC 201604954 (middle panels), and TIC 374095457 (bottom panels), are observed in their wavelet maps and confirmed by FFT and Lomb-Scargle peaks labeled {\em A}.} 
	\label{pul}
\end{figure}

As a by-product of our analysis, the present study has also revealed 10 TOIs with pulsation signatures, with periodicities ranging from 0.049 to 2.995 days. Figure \ref{pul} displays three examples of these pulsating TOIs. A more detailed study would be needed to confirm specific classes of pulsators, a subject that is beyond the scope of the present paper. Table \ref{pulsation} lists these stars with the respective pulsating periods. We have also identified four eclipsing binaries, TIC 9727392, TIC 100100827, TIC 149010208, and TIC 432549364, with orbital periods of 4.534, 0.942, 3.437, and 1.217 days, respectively; the star TIC 149010208 also exhibits two clear flares at September 3rd and 18th 2018.

\begin{deluxetable*}{ccccccccc}
\tablenum{5}
\tablecaption{TOI stars with unambiguous pulsation periodicity from our analysis.\label{pulsation}}
\tablewidth{2pt}
\tablehead{
\colhead{TIC ID} & \colhead{T$_{eff}$} & \colhead{$\log g$} & \colhead{P$_{orb}$} &
\colhead{P$_{pul}$} &\colhead{{\em e}P$_{pul}$} & \colhead{t$_{SPAN}$} & \colhead{N$_{Cycle}$} & \colhead{Sectors} \\
\colhead{}	&	\colhead{(K)}	&	\colhead{(cm/s$^2$)}	&	\colhead{(days)}	&	\colhead{(days)}	&	\colhead{(days)}	&	\colhead{(days)}&	\colhead{}
}
\startdata
129979528	&	7399	&	4.250	&	1.220	&	0.049	&	0.001	&	17	&	346.9	&	 18 \\      
149833117	&	6578	&	4.310	&	4.052	&	0.303	&	0.002	&	23	&	75.4	&	 20 \\
156987351	&	7691	&	4.117	&	3.063	&	0.082	&	0.001	&	39	&	475.6	&	 6 \\
164173105	&	7164	&	4.257	&	3.073	&	0.572	&	0.008	&	20	&	35.9	&	 16 \\
201604954	&	5760	&	4.393	&	4.606	&	0.629	&	0.008	&	25	&	40.1	&	 13 \\
287196418	&	7106	&	4.200	&	3.695	&	1.016	&	0.009	&	61	&	60.2	&	 14,16,17 \\
297967252	&	8599	&		&	9.683	&	1.128	&	0.014	&	45	&	39.5	&	 9 \\
329277372	&	5780	&	4.438	&	2.888	&	0.524	&	0.003	&	40	&	77.4	&	 16 \\
350132371	&	5811	&	4.130	&	1.032	&	1.990	&	0.035	&	56	&	28.2	&	 16,17,18 \\
374095457	&	5780	&	4.438	&	0.784	&	2.995	&	0.125	&	36	&	11.9	&	 10 9 \\
\enddata
\tablecomments{The following information is listed: the TIC ID, effective temperature (T$_{eff}$), surface gravity ($\log g$), and orbital period (P$_{orb}$) taken from the TOI Release Portal (\url{https://tess.mit.edu/toi-releases/}), pulsation period (P$_{pul}$), error in the pulsation period ({\em e}P$_{pul}$), effective time span (t$_{SPAN}$), and effective number of cycles (N$_{Cycle}$), obtained from our analysis, and TESS observation sectors. Values for $\log g$ and P$_{orb}$ are rounded to 3 decimals digits. }
\end{deluxetable*}

\subsection{KELT periodicities for TOI stars} \label{KELT}

The Kilodegree Extremely Little Telescope (KELT) project (Pepper et al. 2007; 2012), has been surveying bright stars with a typical cadence between 10--30 minutes, for more than four million sources with apparent visual magnitudes in the approximate range 7 $<$ V $<$13.  Dedicated to the search of  transiting of large-radii planets, KELT has also supported studies on the variability of thousands of stars. Oelkers et al. (2018) provided a catalog of 62,229 stars presenting significant large-amplitude fluctuations probably caused by stellar rotation. Indeed, this survey provides rotation periods for a significant amount of stars in common with the TESS catalog, using a homogeneous procedure, offering the possibility of a comparison with periods obtained from the present study. Fourty objects of our present sample of TOIs are listed by those authors as stars with likely rotation periods. 

For seventeen of the referred stars we have identified only noise in their TESS LCs, whereas the eighteen additional stars have confirmed rotation periods. As shown in Table \ref{kelt1}, for this second group, the rotation periods for 9 stars are in agreement, within a range of 10\%, and the other 9 stars are in disagreement, when comparing our period measurements with those by Oelkers et al. (2018). Table \ref{kelt2} lists the sub-sample of TOI stars with noisy LCs, with the periodicities computed by Oelkers et al. (2018) ranging from about  0.9 to 47 days. A comparison between rotation periods measured in the present study and those estimated by Oelkers et al. (2018) should be taken with cautious because different aspects are involved in the observational procedures, including  the cadence of observations, observational time spans and, in particular, the photometric precisions. Nevertheless, it is worthy to underline that for the noisy stars from TESS, namely those stars for which we have found no periodicities, Oelkers et al. (2018) were able to estimate periods for 12 stars with values larger than 14 days. For these noisy stars low activity or long rotation periods are expected, which is fairly in parallel with the high periods found by Oelkers et al. (2018). 

\begin{deluxetable*}{ccc}
\tablenum{6}
\tablecaption{TOI stars with unambiguous rotation from our analysis in common with the KELT catalog of rotation periodicity (Oelkers et al. 2018).\label{kelt1}}
\tablewidth{2pt}
\tablehead{
\colhead{TIC ID} & \colhead{P$_{rot}$} &
\colhead{P$_{rot}$} \\
\colhead{}	&	\colhead{Our work}	&\colhead{KELT}\\
\colhead{}	&	\colhead{(days)}	&\colhead{(days)}
}
\startdata
9348006	&	5.329	&	15.7332	\\
13499636	&	5.595	&	1.12709	\\
22843856	&	3.518	&	3.8088	\\
29191596	&	9.070	&	9.88338	\\
138017750	&	2.623	&	1.23605	\\
153949511	&	8.095	&	27.1518	\\
156991337	&	3.607	&	3.76619	\\
201248411	&	13.219	&	12.7535	\\
207141131	&	8.490	&	8.69263	\\
219776325	&	9.330	&	10.0806	\\
220459826	&	5.559 & 0.521154	\\
229938290	&	8.600	&	1.30302	\\
235037761	&	7.359	&	7.30887	\\
241196395	&	2.019	&	1.04978	\\
293954617	&	5.368	&	11.7178	\\
356311210	&	5.356	&	5.36711	\\
382474101	&	2.722	&	0.728157	\\
459970307	&	3.581	&	7.05368	\\
\enddata
\end{deluxetable*}

\begin{deluxetable*}{cc}
\tablenum{7}
\tablecaption{TOI stars with a noisy behavior from our analysis in common with the KELT catalog of rotation periodicity (Oelkers et al. 2018).\label{kelt2}}
\tablewidth{2pt}
\tablehead{
\colhead{TIC ID} & \colhead{P$_{rot}$ (KELT)} \\
	&	\colhead{(days)}	
}
\startdata
69679391	&	29.5334	\\
115771549	&	35.8166	\\
130924120	&	14.1864	\\
134200185	&	45.4959	\\
167754523	&	17.6585	\\
207084429	&	14.1864	\\
237928815	&	21.3995	\\
257241363	&	32.3729	\\
279741379	&	47.6417	\\
286355915	&	0.961816	\\
306996324	&	1.04328	\\
309792357	&	1.04328	\\
322063810	&	0.902519	\\
377293776	&	25.1256	\\
403224672	&	1.12583	\\
406672232	&	30.4229	\\
413248763	&	20.5297	\\
\enddata
\end{deluxetable*}
\section{Summary} \label{summary}

We conduct an in-depth search for rotation and pulsation signatures from a sample of 1000 TOI stars observed in 2-min cadence by TESS. Such an analysis was based on three procedures, namely wavelet, Fast Fourier Transform and Lomb-Scargle, along with a meticulous visual inspection. We identified 163 TOIs with clear rotational modulation, from which 131 stars present unambiguous rotation period, ranging from 0.321 to 13.219 days, one of these stars being fast rotator with P$_{rot}~<$~0.50 days, and 32 of them presenting dubious values for the periodicity. The present analysis revealed also four eclipsing binaries, ten stars presenting clear signatures of pulsation, with periods ranging from 0.049 to 2.995 days, and 109 stars show ambiguous variability, whose astrophysical root-cause is not clearly identified. For the remaining 714 TOIs the TESS light curves show essentially a noisy pattern, with low amplitude signals.  Whereas the signatures of rotation reflect the presence of prominent star spots at different locations in the stellar surface, the stars with ambiguous variability and noisy pattern appear to reflect a large number of causes, including polar spots, low activity phases and long periodicity. In this sense, among the 17 stars with TESS LCs presenting noisy pattern, in common with KELT observations, 12 have KELT periods ranging from 14 to 47 days, therefore rotating slower than our sample with unambiguous rotation periodicities. The scenario for rotation from an analysis combining the present results with those from Oelkers et al. (2018) tend to follow fairly the same trend observed by different authors (e.g.: Le\~ao et al. 2015; Paz-Chinch\'on et al. 2015; McQuillan et al. 2013), in particular for {\em Kepler} stars with planet candidates. As reported by those authors, rotation period for M to F stars are distributed typically from about 1 to 80 days. In addition, studies measuring the rotation periods of nearby low-mass stars (Newton et al. 2016, 2018) have identified a population of fast rotators (P$_{rot} <$ 10 d) and a population of slow rotators (P$_{rot} >$ 70 d), a fact that is followed by the present results considering only the group of fast rotators found by those authors.

We shall also touch upon for some particularities emerging from the present analysis: 22 stars have P$_{rot}\simeq$ P$_{orb}$. Within the observational uncertainties, this finding points for potential targets undergoing a stage of tidal synchronization. This study revealed also 23 TOIs with unambiguous rotation period showing two periods, one being approximately the double or a half of the rotation period, as it can be clearly seen in the wavelet maps (e.g., Fig. \ref{rot}, middle panel). As mentioned in Sects. 2.1 and 2.4 and as described in Basri \& Nguyen (2018), this is a common pattern observed in rotating stars that is overall related to hemispherical asymmetries. Although those asymmetries may be associated to a complex spot distribution and dynamics, they can be explored with the help of relatively simple spot modeling, providing thus important clues for the study of spot dynamics and differential rotation (e.g.: Lanza et al. 2014; Aigrain et al. 2015; das Chagas et al. 2016). Another particular aspect regards to the group of ten stars with pulsation, which offers additional perspectives to explore the frequency modulation methods (Shibahashi \& Kurtz 2012), to derive information traditionally obtained from radial velocity procedure (e.g.: Murphy et al. 2014, 2016; Hermes 2018).

Let us also underline that all the types of stellar variability, as source of identified astrophysical phenomena or noise, can impact directly on the precision and accuracy of exoplanet multi-band photometric transit and spectroscopic observations. Fast rotators, particularly, can inhibit the detection of small planets (e.g.: Berta et al. 2012; Kipping et al. 2017), whereas measurements of planetary mass can be hindered if a star and its planet are in tidal synchronization. In this sense, the results pointed out in this paper offer also constraints to predict the impact of stellar variability resulting from rotation and pulsation on observations dedicated to the characterization of planets around TOI stars, and present a methodology that could be applied to other samples of stars with or without planetary companions. As TESS continues to observe the sky, it will produce a large quantity of 2-min cadence LCs for thousands of stars, an additional unique laboratory for the continuation of the work reported here. 

Finally, this study reinforces an important lesson: for identifying periodicities with real physical meaning, it is not sufficient the selection of numbers emerged from periodograms obtained from a single computational method. In many occasions, such periods may be mere artifacts of the method used or may represent only estimations. In this context, obtaining periodicities based on multiple methods that combine information from different types of periodograms together with wavelet analysis, which provides the identification of periodicity associated with the persistence of the phenomenon, as well as with a visual inspection of the LC, is the recommended path for a more confident determination of periodicities with clear astrophysical meaning.

The full catalog has been uploaded at the Filtergraph portal\footnote{\url{https://filtergraph.com/tess_rotation_tois }} (Burger et al. 2013) for data visualization. The portal can be used to access the variability and periodicity information described in this study: stellar parameters obtained from the TESS data basis, LC, FFT and Lomb-Scargle periodograms, and wavelet maps for each TOI star. This portal is meant to be a living database and will be updated with new rotation and pulsation periods as soon as new TOIs LCs become public at the TESS portal.   

\acknowledgments

We warmly thank our families for involving us with care, patience, and tenderness, during the home office tasks for the preparation of this work face to this COVID-19 difficult moment. Research activities of the observational astronomy board at the Federal University of Rio Grande do Norte are supported by continuous grants from the brazilian funding agencies CNPq, FAPERN, and INCT-INEspaço. This study was financed in part by the Coordenação de Aperfeiçoamento de Pessoal de Nível Superior - Brasil (CAPES) - Finance Code 001. RLG, YSM, and MAT acknowledge CAPES graduate fellowships, and ABB acknowledges CNPq graduate fellowship. JPB acknowledges a CAPES/PNPD fellowship. ICL, BLCM, and JRM acknowledge CNPq research fellowships. This paper includes data collected by the TESS mission. Funding for the TESS mission is provided by the NASA Explorer Program. The research described in this paper makes use of Filtergraph, an online data visualization tool developed at Vanderbilt University through the Vanderbilt Initiative in Data-intensive Astrophysics (VIDA) and the Frist Center for Autism and Innovation (FCAI). We warmly thank the Referee for comments and suggestions that clarified important aspects of this study.
    


%





\end{document}